\documentclass[sigconf,screen]{acmart}
\AtBeginDocument{%
  \providecommand\BibTeX{{%
    Bib\TeX}}}

\usepackage{listings}
\usepackage{xcolor}
\usepackage{xurl} 
\usepackage{hyperref}
\usepackage{tikz}
\usepackage{standalone}
\usetikzlibrary{shapes, arrows.meta, positioning, fit, backgrounds, calc, shadows}
\usepackage{adjustbox}

\def\BibTeX{{\rm B\kern-.05em{\sc i\kern-.025em b}\kern-.08em
    T\kern-.1667em\lower.7ex\hbox{E}\kern-.125emX}}

\setcopyright{acmlicensed}
\copyrightyear{2026}
\acmYear{2026}
\acmDOI{XXXXXXX.XXXXXXX}
\acmConference[Conference acronym 'XX]{Make sure to enter the correct
  conference title from your rights confirmation email}{June 03--05,
  2026}{Woodstock, NY}

\begin{document}

\title{Auditing MCP Servers for Over-Privileged Tool Capabilities}



\author{Charoes Huang}
\affiliation{
  \institution{Department of Computer Science}
  \institution{New York Institute of Technology}
  \city{Vancouver}
  \state{BC}
  \country{Canada}}
\email{yhuang93@nyit.edu}

\author{Xin Huang}
\affiliation{
  \institution{Department of Computer Science}
  \institution{New York Institute of Technology}
  \city{Vancouver}
  \state{BC}
  \country{Canada}}
\email{xhuang31@nyit.edu}

\author{Amin Milani Fard}
\affiliation{
  \institution{Department of Computer Science}
  \institution{New York Institute of Technology}
  \city{Vancouver}
  \state{BC}
  \country{Canada}}
\email{amilanif@nyit.edu}


\begin{abstract}
The Model Context Protocol (MCP) has emerged as a standard for connecting Large Language Models (LLMs) to external tools and data. However, MCP servers often expose privileged capabilities, such as file system access, network requests, and command execution that can be exploited if not properly secured. We present \texttt{mcp-sec-audit}, an extensible security assessment toolkit designed specifically for MCP servers. It implements static pattern matching for Python-based MCP servers and dynamic sandboxed fuzzing and monitoring via Docker and eBPF. The tool detects risky capabilities through configurable rule-based analysis and provides mitigation recommendations.
\end{abstract}




\maketitle

\section{Introduction}

As LLM agents increasingly rely on external tools via protocols such as MCP~\cite{anthropic2024mcp}, the security boundary shifts from the model to the servers providing these tools \cite{mcp-sec-analysis}. Existing Static Application Security Testing (SAST) tools do not model MCP-specific artifacts such as tool metadata/descriptions, and therefore cannot directly report exposed capabilities from MCP tool definitions. There is a lack of dedicated tooling to audit MCP servers for security risks before they are deployed in sensitive environments. Traditional SAST tools identify risky code patterns such as command execution and insecure file handling; however, they operate on source code alone and lack protocol-specific awareness. 

We present \texttt{mcp-sec-audit}\footnote{\url{https://github.com/nyit-vancouver/mcp-sec-audit}}, which analyzes both the implementation code and MCP tool metadata, and reports capability-based deployment risks with actionable hardening guidance. Existing vulnerability scanners focus on package dependencies and container misconfigurations. While complementary, they do not infer runtime capabilities from tool definitions. Our tool bridges this gap by mapping code-level indicators to capability categories such as \texttt{command\_exec} and \texttt{file\_write}, and pairing them with least-privilege deployment recommendations. Many defenses focus on runtime policy enforcement. \texttt{mcp-sec-audit} operates earlier in the lifecycle as a pre-deployment audit tool using rule-driven static analysis and optional sandbox-based dynamic verification.

\textbf{Contributions.} This work makes the following contributions:

\begin{itemize}
    \item \textbf{Auditing Framework:} We present \texttt{mcp-sec-audit}, an auditing framework that automatically identifies high-risk capabilities in MCP servers and outputs deployment-oriented mitigation guidance such as least-privilege container and filesystem recommendations. 
    \item \textbf{Modular Pipeline Architecture:} A pipeline-based architecture with a static pipeline (TOML-driven keyword/regex matching over Python and metadata), and a dynamic pipeline that supports sandbox execution and telemetry collection on eBPF-enabled Linux hosts.
    \item \textbf{Protocol-Aware Detection:} Protocol-aware static detectors for (1) Python source code and (2) MCP tool metadata/descriptions, driven by a unified TOML rulebook that defines capability families, indicators, and severity weights.
    \item \textbf{Mitigation Recommendation Engine:} Mapping detected capabilities such as \texttt{file\_write} to deployment hardening recommendations using a built-in mapping with a placeholder interface for loading external templates.
    \item \textbf{Extensible Plugin System:} A lightweight registry abstraction that enables adding new detectors without modifying the pipeline logic.
\end{itemize}

\begin{figure*}[t]
    \centering
    \begin{adjustbox}{max width=0.8\textwidth}
        \begin{tikzpicture}[
    node distance=1cm and 1.5cm,
    component/.style={
        rectangle, draw=black!70, thick,
        fill=white,
        minimum height=1.2cm, minimum width=2.8cm,
        align=center, font=\small\sffamily,
        rounded corners=3pt,
        drop shadow={opacity=0.25, shadow xshift=2pt, shadow yshift=-2pt}
    },
    engine/.style={
        component,
        fill=blue!5, draw=blue!60, thick, font=\small\sffamily\bfseries
    },
    store/.style={
        cylinder, shape border rotate=90, draw=black!70, thick,
        fill=gray!10, aspect=0.25,
        minimum height=1.5cm, minimum width=3.0cm,
        align=center, font=\footnotesize\sffamily
    },
    entity/.style={
        rectangle, draw=none,
        font=\small\sffamily\bfseries,
        align=center 
    },
    groupbox/.style={
        draw=gray!30, dashed, fill=gray!3,
        inner sep=0.6cm, rounded corners=8pt,
        label={[anchor=north west, text=gray!70, font=\bfseries\scriptsize, xshift=5pt, yshift=-5pt]north west:\textsc{#1}}
    },
    conn/.style={-{Latex[length=3mm]}, thick, draw=black!60, rounded corners=10pt}
]

    \node[entity] (user) {User / CLI};
    
    \node[component,right=of user, align=left] (orchestrator) {
        \textbf{Orchestration Core}\\
        (Pipeline Manager,\\
        Session Context)
    };

    \node[store, below=0.5cm of user] (rules) {Security\\Rules \&\\Policies};
    \node[store, below=1.5cm of orchestrator] (models) {Data Models\\(Findings)};

    
    \node[engine, right=2.5cm of orchestrator] (static_eng) {Static Analysis\\Engine};
    
    \node[component, below=0.3cm of static_eng, minimum height=0.8cm, font=\footnotesize] (ast) {Pattern \& Metadata\\Analyzers};

    \node[engine, below=1.5cm of static_eng] (dynamic_eng) {Dynamic Analysis\\Engine};
    \node[component, below=0.3cm of dynamic_eng, minimum height=0.8cm, font=\footnotesize, fill=red!5, draw=red!50] (sandbox) {Sandbox /\textbf{eBPF Monitor}};

    \coordinate (right_col_x) at ($(static_eng.east) + (3.5cm, 0)$);
    
    \node[component] (synthesis) at (right_col_x |- static_eng) {
        \textbf{Risk Scoring}\\
        (Aggregate Findings,\\
        Calculate Score)
    };
    
    \node[component, below=1.2cm of synthesis] (mitigation) {Mitigation\\Generator};
    \node[entity, below=1.0cm of mitigation] (report) {Final Security\\Report};

    \coordinate (web_col_x) at ($(synthesis.east) + (2.5cm, 0)$);

    \node[component] (frontend) at (web_col_x |- synthesis) {Frontend\\Interface};
    \node[component, below=1.2cm of frontend] (api) {RESTful API\\Layer};

    \draw[conn] (user) -- (orchestrator);
    \draw[conn] (orchestrator.east) -- (static_eng.west);
    \draw[conn] (orchestrator.east) -- ++(0.5,0) |- (dynamic_eng.west);
    
    \draw[conn, dashed] (rules) -- (orchestrator);
    \draw[conn, dashed] (rules.east) -- (ast.west);
    
    
    \path (ast.south) -- (dynamic_eng.north) coordinate[midway] (middle_gap_y);
    \draw[conn, dashed] (models.north) |- (middle_gap_y) -| (synthesis.west);

    \draw[thick, draw=blue!40] (static_eng) -- (ast);
    \draw[thick, draw=red!40] (dynamic_eng) -- (sandbox);
    
    \draw[conn] (static_eng) -- (synthesis);

    \draw[conn] (dynamic_eng.east) -- ++(1.5,0) |- (synthesis.south west);

    \draw[conn] (synthesis) -- (mitigation);
    \draw[conn] (mitigation) -- (report);

    \draw[conn] (report.east) -| (api.south);
    \draw[conn] (api) -- (frontend);

    \begin{scope}[on background layer]
        \node[groupbox={Control Plane \& Knowledge Base}, fit=(user) (orchestrator) (rules) (models)] {};
        \node[groupbox={Core Analysis Plane}, fit=(static_eng) (ast) (dynamic_eng) (sandbox)] {};
        \node[groupbox={Output Generation Plane}, fit=(synthesis) (mitigation) (report)] {};
        \node[groupbox={Web Detection Portal}, fit=(frontend) (api)] {};
    \end{scope}

\end{tikzpicture}
    \end{adjustbox}
    \caption{System Architecture of the MCP Security Analyzer.}
    \label{fig:arch}
\end{figure*}
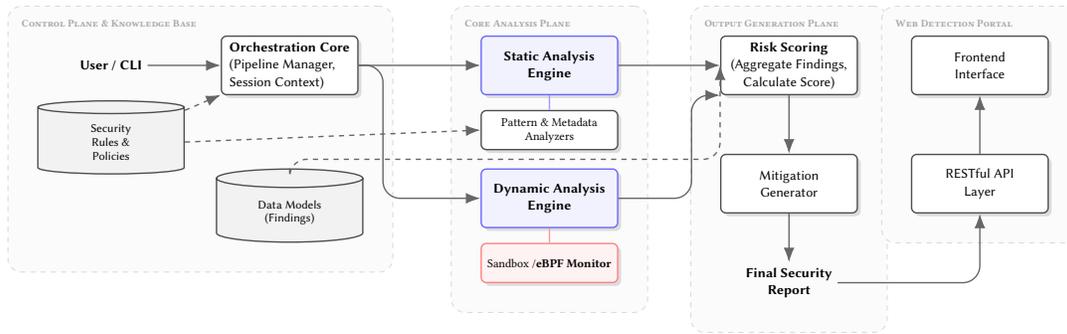

\begin{figure}[t]
    \includegraphics[trim = 5mm 40mm 0mm 0mm,width=0.91\textwidth]{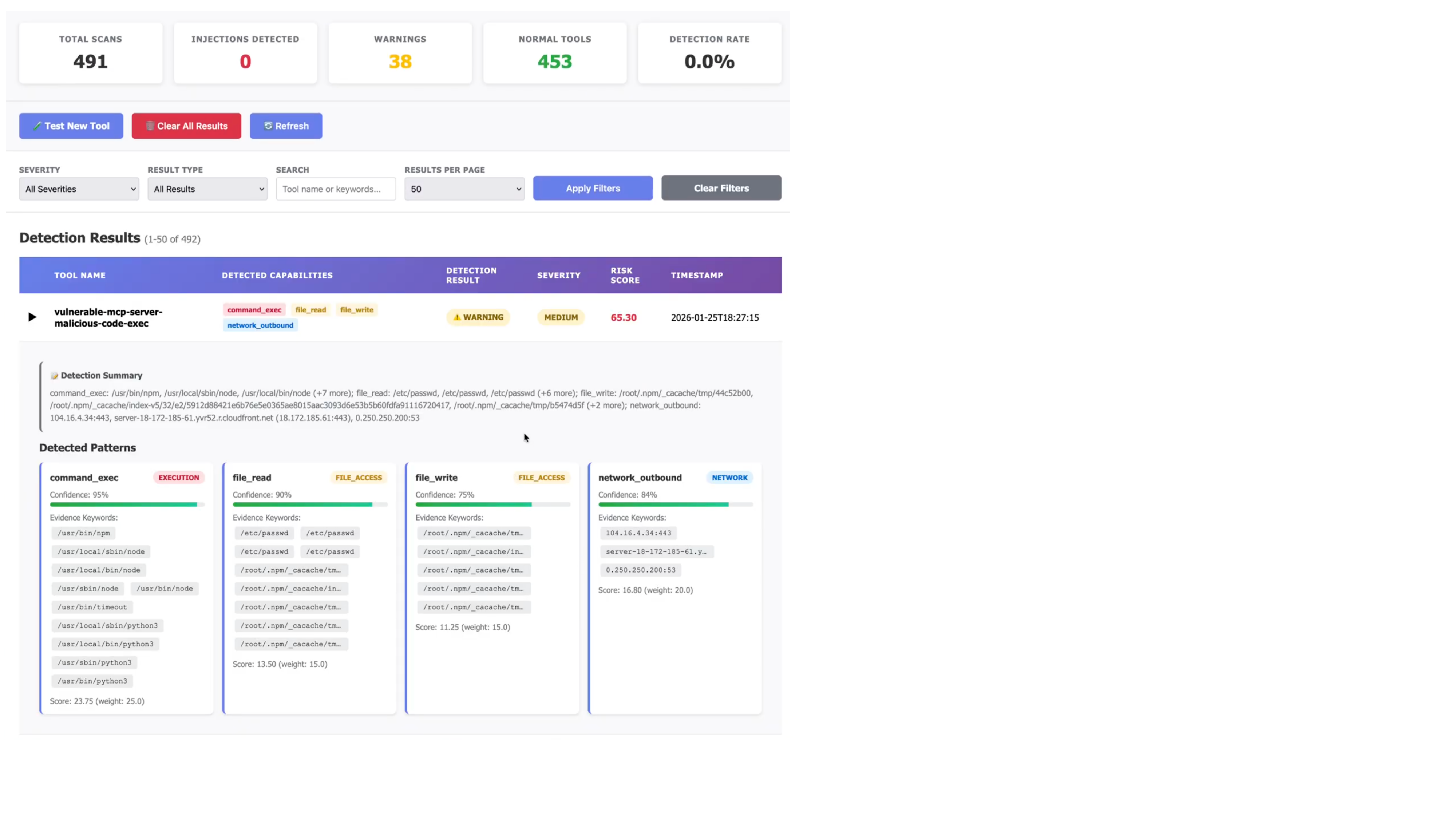}
    \caption{A Screenshot of the Web-based Detection Portal.}
    \label{screenshot}
\end{figure}

Our tool provides pre‑deployment auditing of MCP servers, while several runtime defenses offer complementary protection by filtering prompts and model outputs during execution. Azure Prompt Shields detect prompt‑injection attempts in real time \cite{azure2025promptshields}, Llama Guard 3 performs safety classification for both inputs and responses \cite{grattafiori2024llama}, and LLM‑Guard supplies prompt/output scanners for production systems \cite{llmguard_protectai}. Complementing these layers, assessing security issues by simulating action results and applying LLM-based safety evaluators \cite{Ruan24,lin2025actionsafetyeval} can help users identify the security severity of actions performed by agents.

\section{Tool Architecture and Implementation}

We designed our tool based on a modular architecture that is shown in Figure \ref{fig:arch} and is described as follows:

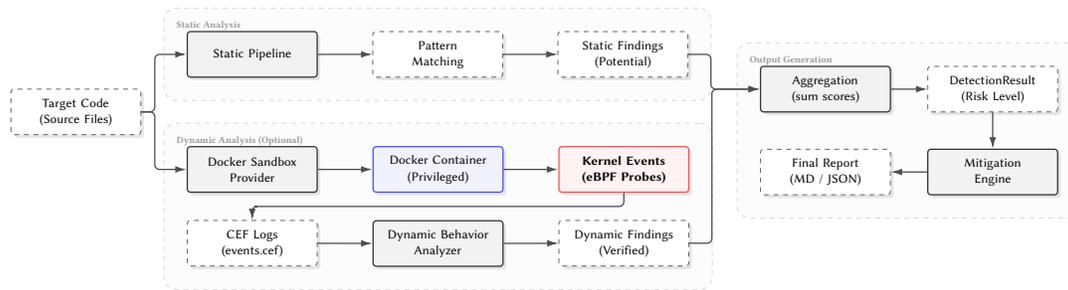
\begin{figure*}[ht]
    \centering
    \begin{adjustbox}{max width=0.8\linewidth}
        \begin{tikzpicture}[
    node distance=0.8cm and 1.2cm, 
    base/.style={
        rectangle, 
        rounded corners=2pt, 
        minimum height=1.0cm,      
        text width=2.5cm,          
        align=center, 
        font=\sffamily\small,
        draw=black!80,
        fill=white,
        drop shadow={opacity=0.2}, 
        inner sep=5pt
    },
    process/.style={base, fill=gray!10, thick},
    data/.style={base, dashed, draw=black!60},
    container/.style={base, draw=blue!70, thick, fill=blue!5},
    ebpf/.style={base, draw=red!70, thick, fill=red!5, font=\sffamily\bfseries\small},
    group/.style={
        draw=gray!30, 
        dashed, 
        inner sep=0.5cm, 
        rounded corners=8pt, 
        fill=gray!2,
        label={[anchor=north west, font=\bfseries\scriptsize, text=gray!80, xshift=5pt, yshift=-5pt]north west:#1}
    },
    arrow/.style={-{Latex[length=2.5mm, width=1.5mm]}, thick, black!70, rounded corners=3pt}
]

    \node[process] (static_pipe) {Static Pipeline};
    \node[process, below=1.5cm of static_pipe] (sandbox) {Docker Sandbox\\Provider};
    
    \path (static_pipe.west) -- (sandbox.west) coordinate[midway] (midpoint_input);
    \node[data, left=1.0cm of midpoint_input] (code) {Target Code\\(Source Files)};

    
    \node[data, right=of static_pipe] (ast) {Pattern\\Matching};
    \node[data, right=of ast] (static_find) {Static Findings\\(Potential)};

    \node[container, right=of sandbox] (container) {Docker Container\\(Privileged)};
    \node[ebpf, right=of container] (kernel) {Kernel Events\\(eBPF Probes)};
    
    \node[data, below=0.6cm of sandbox] (cef) {CEF Logs\\(events.cef)};
    \node[process, below=0.6cm of container] (analyzer) {Dynamic Behavior\\Analyzer};
    \node[data, below=0.6cm of kernel] (dyn_find) {Dynamic Findings\\(Verified)};

    
    \path (static_find.east) -- (kernel.east) coordinate[midway] (midpoint_right);
    \coordinate (matrix_center) at ($(midpoint_right) + (3.0cm, 0.5cm)$); 
    
    \node[process] (merger) at (matrix_center) {Aggregation\\(sum scores)};
    
    \node[data, right=0.8cm of merger] (result) {DetectionResult\\(Risk Level)};
    
    \node[process, below=0.8cm of result] (mit_engine) {Mitigation\\Engine};
    
    \node[data, left=0.8cm of mit_engine] (report) {Final Report\\(MD / JSON)};

    \draw[arrow] (code.east) -- ++(0.3,0) |- (static_pipe);
    \draw[arrow] (code.east) -- ++(0.3,0) |- (sandbox);

    \draw[arrow] (static_pipe) -- (ast);
    \draw[arrow] (ast) -- (static_find);

    \draw[arrow] (sandbox) -- (container);
    \draw[arrow] (container) -- (kernel);
    \draw[arrow] (kernel.south) -- ++(0,-0.3) -- ++ (-8,0) -|(cef.north);
    \draw[arrow] (cef) -- (analyzer);
    \draw[arrow] (analyzer) -- (dyn_find);

    \draw[arrow] (static_find.east) -- ++(0.5,0) |- (merger.west);
    \draw[arrow] (dyn_find.east) -- ++(0.5,0) |- (merger.west);

    \draw[arrow] (merger) -- (result);
    \draw[arrow] (result) -- (mit_engine);
    \draw[arrow] (mit_engine) -- (report);

    \begin{scope}[on background layer]
        \node[group={Static Analysis}, fit=(static_pipe) (ast) (static_find)] {};
        \node[group={Dynamic Analysis (Optional)}, fit=(sandbox) (container) (kernel) (cef) (analyzer) (dyn_find)] {};
        \node[group={Output Generation}, fit=(merger) (result) (mit_engine) (report)] {};
    \end{scope}

\end{tikzpicture}
    \end{adjustbox}
    \caption{Runtime Workflow of Dual-stream Analysis with Static Inspection and Dynamic eBPF Monitoring.}
    \label{fig:data_flow}
\end{figure*}

\begin{enumerate}
    \item \textbf{Control Plane and Knowledge Base}
    \begin{itemize}
        \item \textbf{Orchestration Core:} This component serves as the system's entry point. It initializes the \texttt{DetectionSession} to maintain runtime state and manages the lifecycle of plugins via the \texttt{DetectorRegistry}. It orchestrates the flow of data between the user CLI and the analysis engines.
        \item \textbf{Security Rules and Policies:} A TOML-based rulebook such as \texttt{detection/rules/keywords.toml} defines capability families, high-risk indicators (keywords/regex), and severity weights, serving as the knowledge base for both analysis streams.
    \end{itemize}

    \item \textbf{Core Analysis Plane}
    \begin{itemize}
        \item \textbf{Static Analysis Engine:} Scans source code and processes metadata configuration files (JSON). This engine utilizes efficient pattern matching with regex and keywords to identify potential capabilities by validating against the loaded rulebook.
        \item \textbf{Dynamic Analysis Engine:} Executes the target tool within an isolated Docker Sandbox. An active Protocol Fuzzer stimulates tool endpoints with malicious payloads, while a kernel-level eBPF Monitor attaches to the container to capture runtime telemetry (syscalls, file I/O, network connections), serializing these events into Common Event Format (CEF) logs for behavioral reconstruction.
    \end{itemize}

    \item \textbf{Output Synthesis Plane}
    \begin{itemize}
        \item \textbf{Risk Scoring:} The pipeline calculates the total risk score as a weighted sum of capability confidences and maps it to a risk level (low/medium/high/critical).
        \item \textbf{Mitigation Generator:} Maps detected capabilities to  deployment hardening recommendations such as Docker isolation, read-only mounts, and network restrictions.
        \item \textbf{Report Builder:} Renders results in a MD or JSON format.
    \end{itemize}

\item \textbf{Web Detection Portal}
\begin{itemize}
    \item \textbf{Front-end Interface Layer:} An interactive web dashboard parses and visualizes JSON results in sortable, filterable tables with columns for severity, result type, and detected keywords. A sample screenshot is shown in Figure \ref{screenshot}. Risk score is visualized using color-coded indicators based on configurable thresholds with expandable rows showing detailed capability breakdowns and traces. 
    
    \item \textbf{RESTful API Layer:} 
    \begin{itemize}
        \item \texttt{POST /api/detect} - Primary detection endpoint accepting MCP tool description for real-time detection and returning severity based on keyword matching.
        \item \texttt{GET /api/results} - Returns the complete detection result as JSON including identified capabilities, risk scores, and evidence chains.
        \item \texttt{GET /api/stats} - Returns aggregated statistics including detector configuration (categories: file\_access, network, execution), risk thresholds, total keyword rules, and storage statistics (injection count, warning count, normal count, total scans, injection rate).
        \item \texttt{POST /api/import/detection} - Batch import endpoint for pre-computed detection results from external scanners, updating the storage statistics and enabling trend analysis.
    \end{itemize}
\end{itemize}
\end{enumerate}

\textbf{Runtime Workflow.} As illustrated in Figure~\ref{fig:data_flow}, the system processes the target tool through the following steps:
\begin{enumerate}
    \item \textbf{Initialization:} The user invokes the CLI with a target directory. The system initializes a \texttt{DetectionSession} and loads the security rulebook into the \texttt{DetectorRegistry}.
    \item \textbf{Static Analysis:} The {Static Pipeline} scans source code and metadata files, using registered detectors to identify potential capabilities and generate initial static findings.
    \item \textbf{Dynamic Fuzzing:} If triggered, the {Docker Sandbox Provider} executes the tool in an isolated container where a protocol-aware fuzzer injects payloads such as command injection patterns. Simultaneously, an eBPF monitor captures kernel-level events (syscalls, I/O) and serializes them into CEF logs.
    \item \textbf{Behavioral Analysis:} The {Dynamic Behavior Analyzer} parses CEF-formatted logs to extract capability evidence (command execution, file I/O, network connections).
    \item \textbf{Scoring:} The pipeline aggregates findings and calculates the total risk score as a weighted sum of detected capabilities.
    \item \textbf{Mitigation:} Finally, the {Mitigation Engine} maps confirmed capabilities to specific hardening actions, and a comprehensive security report is generated.
\end{enumerate}

\textbf{Implementation Details.} We implemented our tool in Python with a plugin-based architecture for extensibility.

\begin{itemize}
    \item \textbf{Static Analysis:} We utilize efficient regular expression and keyword matching to identify patterns of privileged operations such as \texttt{subprocess.run} and \texttt{open(..., 'w')}. Rule definitions are in \texttt{TOML} files (\texttt{rules/\allowbreak keywords.\allowbreak toml}), currently supporting Python source code and MCP metadata JSON files.
    \item \textbf{Dynamic Analysis:} It includes infrastructure for Docker SDK-based container management and CEF log parsing. The engine includes a Python-based MCP Fuzzer that handles the protocol handshake and automatically injects payloads such as shell injection and path traversal into tool arguments. The implementation expects an external Docker image with eBPF instrumentation to generate runtime telemetry.
    \item \textbf{Data Models:} Strictly typed \texttt{dataclasses} define models including \texttt{DetectionResult} and \texttt{Capability\allowbreak Finding}, ensuring consistency across analysis modes.
\end{itemize}

\section{Installation, Demonstration, and Scenarios}

The system requires Python 3.10+ and Docker. The installation steps are explained in the README.md file on our Github repository. For more information about the code, security detection framework, calculations of risk score and confidence, and web-based detection portal, please visit the project page\footnote{\url{https://nyit-vancouver.github.io/mcp-sec-audit/}}. In addition, a video demonstration\footnote{\url{https://tinyurl.com/mcp-sec-audit-demo}} presents the architecture and usage of the tool.

To confirm installation, run the CLI help command. Then test it on a sample server:

\begin{lstlisting}
cd detection 
uv run -m detection.cli --help
uv run -m detection.cli \
    examples/vulnerable-mcp-servers-lab/vulnerable-mcp-server-malicious-code-exec \
    --pipeline dynamic
\end{lstlisting}
The output displays detected capabilities and a risk assessment.

\textbf{Scenario A: Pre-deployment Audit.} A developer has created a new Python-based MCP server \texttt{financial-tools} capable of reading local reports.
    \begin{lstlisting}
    uv run -m detection.cli ./financial-tools --pipeline static --format markdown
    \end{lstlisting}
The tool generates a risk report that flags the \texttt{file\_read} capability as \texttt{HIGH} risk and recommends restricting the mount path.


\textbf{Scenario B: Continuous Integration Gate.} A DevOps engineer wants to prevent over-privileged servers from merging. This is done by configuring the CI pipeline to run the detection tool:
    \begin{lstlisting}
uv run -m detection.cli ./target --pipeline static --format json
    \end{lstlisting}
    The build should fail if \texttt{risk\_level} is \texttt{CRITICAL}. As the output, a Pull Request introducing arbitrary shell execution (\texttt{os.system}) would be flagged, with the JSON report identifying files containing the matched patterns.


\textbf{Scenario C: Mitigation Recommendations.} A security engineer needs guidance on deploying a third-party MCP server. By running the static analyzer, the \texttt{MitigationEngine} provides general recommendations based on detected capabilities, such as "Mount only required directories, preferably read-only" for file access or "Run with no-new-privileges" for command execution.





\section{Evaluation}


We evaluate the detection rate of \texttt{mcp-sec-audit} along three experiments on a controlled vulnerable server, the MCPTox academic dataset \cite{wang2025mcptox}, and a curated vulnerability lab.

\textbf{Static Analysis Assessment.} We test the static analyzer on a synthesized malicious tool (\texttt{examples/static\_analysis\_test.py}) with command execution, file I/O, and network operations. The static pipeline successfully identified all three capability categories with confidence scores ranging from 0.65 to 0.85, assigned a \texttt{MEDIUM} risk level (total score: 42.5/100), and generated 5 mitigation recommendations including Docker isolation and read-only mounts. The analysis completed in under 2 seconds without requiring Docker.




\textbf{MCPTox Benchmark Assessment.} We evaluate the detection capability of our tool on the MCPTox benchmark for tool poisoning attack on 45 real-world MCP servers ~\cite{wang2025mcptox}. Table~\ref{tab:mcptox-results} summarizes the detection results. The tool identified 663 capability instances with 100\% detection rate across 491 samples (1.35 capabilities per sample on average - many tools triggered multiple capability types). The most frequent ones were \texttt{file\_write}, \texttt{tool\_sequence\_hijack}, and \texttt{prompt\_injection}. In total, 367 out of 491 samples were detected, indicating a detection rate of 74.7\%. 124 samples (25.3\%) did not have explicit capability indicators in the dataset metadata, though some may still be detected by broader rules. Note that we get 100\% detection rate when capability indicators are present. Risk distribution is 92.3\% \texttt{LOW} and 7.7\% \texttt{MEDIUM}, with average score of 9.96 and max 49.05.




\begin{table}[t]
\centering
\caption{MCPTox Benchmark Detection Results.}
\label{tab:mcptox-results}
\begin{tabular}{lr}
\hline
\textbf{Capability} & \textbf{Detected} \\
\hline
\texttt{file\_write}                  & 179 \\
\texttt{tool\_sequence\_hijack}       & 136 \\
\texttt{prompt\_injection}            &  87 \\
\texttt{param\_override}              &  60 \\
\texttt{network\_outbound}            &  47 \\
\texttt{command\_exec}                &  46 \\
\texttt{file\_read}                   &  43 \\
\texttt{network\_inbound}             &  41 \\
\texttt{env\_access}                  &  24 \\
\hline
\textbf{Total}               & 663 \\
\hline
\end{tabular}
\end{table}

\textbf{Static and Dynamic Analysis on Vulnerable Lab.} 
We evaluated both static and dynamic pipelines on Vulnerable MCP Servers Lab \footnote{\url{https://github.com/appsecco/vulnerable-mcp-servers-lab}}, a repository containing 9 intentionally vulnerable implementations of MCP servers each implementing a distinct attack vector: RCE via \texttt{eval()}, path traversal, prompt injection, typosquatting, secrets exposure, and dependency vulnerabilities. Each server was analyzed twice: once with static analysis and once with dynamic sandbox execution. Results are shown in Table~\ref{tab:vuln-lab-complete}.

The static analyzer achieved {100\% detection (2/2)} on Python-based servers but {0\% detection (0/7)} on JavaScript-based servers due to current JavaScript/TypeScript AST parsing limitations. The Python-based \texttt{filesystem-workspace-actions} server was flagged as \texttt{MEDIUM} risk (score: 29.0) with detected capabilities \texttt{command\_exec}, \texttt{file\_read}, and \texttt{file\_write}.

In contrast, the dynamic analyzer demonstrated {language-agnostic detection}, achieving {100\% coverage (9/9 servers)} regardless of implementation language. All seven JavaScript servers—previously undetected by static analysis—were successfully identified, with an average risk score of 61.4/100 (classified as \texttt{HIGH} risk). The \texttt{malicious-code-exec} server (containing explicit \texttt{eval()} RCE) received the highest score (65.3), capturing 4 capability categories including command execution with concrete evidence. Dynamic analysis consistently assigned higher scores than static analysis (average increase: +36.2 points), reflecting behavioral evidence from runtime monitoring.

\begin{table}[t]
\centering
\caption{Vulnerable Servers Lab Detection Results.}
\label{tab:vuln-lab-complete}
\small
\begin{tabular}{lccccc}
\hline
\textbf{Server} & \multicolumn{2}{c}{\textbf{Static}} & \multicolumn{2}{c}{\textbf{Dynamic}} & \textbf{$\Delta$} \\
 & Risk & Score & Risk & Score & \\
\hline
\texttt{filesystem-workspace} (Py) & M & 29.0 & M & 48.5 & +19.5 \\
\texttt{wikipedia-streamable} (Py) & L & 15.0 & M & 34.2 & +19.2 \\
\texttt{malicious-code-exec} (JS) & --- & 0.0 & H & 65.3 & +65.3 \\
\texttt{secrets-pii} (JS) & --- & 0.0 & HI & 62.3 & +62.3 \\
\texttt{indirect-prompt-inj} (JS) & --- & 0.0 & H & 60.7 & +60.7 \\
\texttt{indirect-prompt-rmt} (JS) & --- & 0.0 & H & 60.7 & +60.7 \\
\texttt{malicious-tools} (JS) & --- & 0.0 & H & 60.7 & +60.7 \\
\texttt{namespace-typosquat} (JS) & --- & 0.0 & H & 60.7 & +60.7 \\
\texttt{outdated-packages} (JS) & --- & 0.0 & H & 60.7 & +60.7 \\
\hline
\multicolumn{6}{l}{{Detection rates: Static (Py: 2/2, JS: 0/7), Dynamic (Py: 2/2, JS: 7/7)}} \\
\hline
\end{tabular}
\end{table}

\section{Limitations and Future Work}

\textbf{Language Support.} The static analyzer currently supports Python and JSON metadata but not TypeScript/JavaScript, despite rule definitions in \texttt{keywords.toml}. Extending language support requires implementing additional text-based analyzers following the existing plugin interface.

\textbf{Detection Accuracy.} Pattern-based detection is inherently susceptible to false positives (benign code matching risky keywords) and false negatives (obfuscated or indirect invocations). Rule refinement and context-aware heuristics may improve precision. Although the tool identifies prompt injection patterns in tool descriptions, it does not validate runtime behavior against semantic constraints, including parameter manipulation and tool sequence hijacking. Hybrid approaches combining static rules with LLM-based semantic analysis may improve detection accuracy.

\textbf{Dynamic Analysis Deployment.}
The sandbox-based pipeline requires Linux with eBPF support, privileged Docker containers, and manual image builds. Future work includes automating sandbox provisioning and supporting non-Linux environments via alternative instrumentation such as macOS DTrace or Windows ETW.

\textbf{Mitigation Specificity.}
Current recommendations are hardcoded and generic, including directives like "use Docker with no-new-privileges". The tool lacks fine-grained policy templates, notably per-capability seccomp profiles, and does not generate deployment-ready configurations such as Dockerfiles or Kubernetes manifests.

\textbf{Integration and Automation.}
While the CLI enables manual audits, production use cases require CI/CD integration with automated policy enforcement such as blocking PRs with critical-risk findings. We plan to provide GitHub Actions, pre-commit hooks, and SARIF output for SAST platform compatibility.

\bibliographystyle{ACM-Reference-Format}
\bibliography{references}









\end{document}